\documentclass[fp,twocolumn]{jpsj3}
\usepackage{txfonts}
\usepackage{algorithm}
\usepackage{algorithmic}
\usepackage{amsmath}
\usepackage{multirow}
\usepackage{graphicx}

\usepackage{bm}
\usepackage{color}


\title{Making Birth-Death Processes from Backward Fokker-Planck Equations for Computing Expectations in Langevin Systems}

\author{Jun Ohkubo$^{1,2}$}
\inst{$^1$Graduate School of Science and Engineering, Saitama University,
  255 Shimo-Okubo, Sakura-ku, Saitama-shi, 338-8570, Japan\\
  $^2$JST, PREST, 4-1-8 Honcho, Kawaguchi, Saitama 332-0012, Japan} %\\

\abst{
A method to direct evaluation of expectations for Langevin systems (stochastic differential equations) is proposed. The method is based on a birth-death process which is derived using combinations of dummy variables and It{\^o} formula. As a pedagogical example, a double-well system and expectations for sigmoid-type functions are used. It is shown that the proposed method has some merits from computational point of view; only one time-integration for the birth-death process gives expectations for various initial conditions in the original Langevin systems. Furthermore, the same time-integration result is available for computing various center positions of the sigmoid-type functions.
}

\begin{document}
\maketitle

\section{Introduction}
\label{sec_introduction}

Langevin equations or stochastic differential equations are widely used in various research fields such as non-equilibrium physics, biological physics, and financial issues. 
Usually, the Langevin equations are simulated directly using discrete-time approximations such as the Euler-Maruyama approximation; as for the approximations, for example, see Ref.\citen{Kloeden_book}. 
From the direct simulations, we obtain approximated probability density functions of the systems, and various statistical quantities are evaluated from the probability density functions.
However, we sometimes have interests in only a few statistical quantities such as averages, variances.
For this purpose, the (approximate) probability density functions could not be needed.
Furthermore, sometimes one would want to calculate statistics for different initial conditions.
If the direct simulations for the Langevin systems are employed, we should repeat the simulations by using various initial conditions and it might be computationally expensive.

The backward Fokker-Planck equation, which is also known as the Kolmogorov backward equation, is a method to compute a certain statistical quantity directly\cite{Gardiner_book}.
The original Langevin systems and the corresponding partial differential equations (the backward Fokker-Planck equations) are connected through the Feynman-Kac formula\cite{Kac1949}.
Although the original Feynman-Kac formula has been proposed to evaluate a Wiener (or Brownian) functional, limited usage of the Feynman-Kac formula gives directly the backward Fokker-Planck equations\cite{Gardiner_book}.
The solutions of the backward Fokker-Planck equations enable us to obtain expectation values of a certain statistical quantity for \textit{various initial conditions}.
The backward Fokker-Planck equation is solved by deterministic ways using discretization for time and space.

In the present paper, a different method to evaluate statistical quantities directly \textit{using birth-death processes} is proposed.
The birth-death processes are stochastic processes\cite{Gardiner_book}, and there are some numerical methods to solve them;
sometimes we employ numerical time-integration for coupled ordinary differential equations 
(master equations), or sometimes the Monte-Carlo method, so-called the Gillespie algorithm, is used.
To connect the original Langevin systems to the corresponding birth-death processes, some techniques, i.e., introduction of dummy variables, the It{\^o} formula, and function expansions, are employed in the present paper.
Through a pedagogical example, the derivation of the corresponding birth-death processes is demonstrated.
Additionally, numerical checks are performed, which will suggest us some numerical merits of the proposed method.

There are some comments of the current topic. 
The connection between stochastic differential equations and birth-death processes has already been studied as duality in stochastic processes\cite{Giardina2009,Ohkubo2010,Ohkubo2013,Jansen2014,Ohkubo2019}.
As for the duality, for example, see Ref.~\citen{Liggett_book}; the first use of duality by Levy was related to a certain probability defined for Brownian motions, and there are many works related to the duality.
Although recent progress of the duality gives us a systematic method to derive the birth-death processes\cite{Ohkubo2013,Ohkubo2019}, basically the evaluated statistics are limited to moments or correlation functions
\cite{Shiga1986,Doering2003,Giardina2007,Carinci2013}.
The present paper focuses on the calculation of other statistical quantities; the usage of the backward Fokker-Planck equations and the combinations of some techniques proposed in the present paper will give us a simple derivation method to obtain the new direct calculation scheme for statistical quantities.

The paper is organized as follows.
In Sect.~\ref{sec_model}, a model and problem settings are explained; a simple model with a double-well potential is used and expectations for sigmoid-type functions are evaluated.
Section~\ref{sec_derivation} is the main part of the present paper; the corresponding birth-death process is derived using the simple model.
Numerical demonstrations are shown in Sect.~\ref{sec_numerical_results}.
Finally, the concluding remarks and future works are denoted in Sect.~\ref{sec_concluding_remarks}.

\section{Model}
\label{sec_model}

\subsection{Simple one-dimensional model with a double-well potential}

To explain and demonstrate the proposed framework, a simple one-dimensional model with a double-well potential is used here.
The simple example is employed from some pedagogical reasons; the example is related to an escape problem in classical stochastic processes.
Additionally, as we will see later, not only a simple power-expansion but also the Legendre polynomials appear.
It is straightforward to apply the proposed method to other function expansions.

\begin{figure}
\begin{center}
\includegraphics[width=60mm]{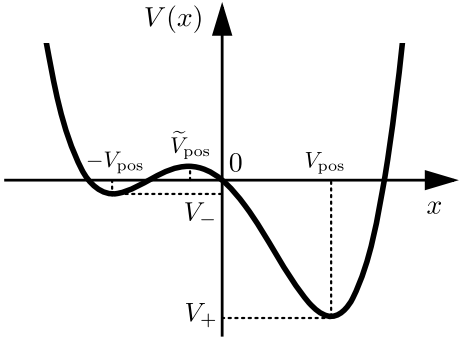}\\
\vspace{5mm}
\end{center}
\caption{A double-well potential used in the simple model.
$x = V_\mathrm{pos}$ and $x = - V_\mathrm{pos}$ give local or global minima.
$V_{+}$ and $V_{-}$ correspond to the potential values at $x = V_\mathrm{pos}$ and $x = - V_\mathrm{pos}$, respectively.
These four (essentially, three) parameters determine the shape of the potential; $\widetilde{V}_\mathrm{pos}$, at which the first derivative of $V(x)$ is zero, is calculated from these four parameters.}
\label{fig_potential} 
\end{figure}

In the following sections, the notation based on the stochastic differential equation is employed\cite{Gardiner_book}.
Here, the following simple stochastic differential equation is considered:
\begin{align}
dx = - \partial_x V(x) \, dt + D \, dW, 
\label{eq_original_sde}
\end{align}
where $V(x)$ is a potential of the system, $D$ is a state-independent diffusion constant,
and $dW$ is the derivative of a conventional Wiener process $W(t)$.
The potential $V(x)$ has the following form;
\begin{align}
V(x) = c_1 x + c_2 x^2 + c_3 x^3 + c_4 x^4.
\label{eq_potential}
\end{align}
That is, the particle moves on a double-well potential system.
Figure~\ref{fig_potential} shows the shape of the double-well potential.
The following four parameters are used to determine the shape of the potential;
$V_\mathrm{pos}$ and $- V_\mathrm{pos}$ give positions of two minima of the system;
$V_{+}$ and $V_{-}$ give the potential values at $x = V_\mathrm{pos}$ and $x = -V_\mathrm{pos}$, respectively.
There is another special point at which the derivative of the potential gives zero;
we denote the $x$ value as $\widetilde{V}_\mathrm{pos}$.
The value of $\widetilde{V}_\mathrm{pos}$ is calculated from the above four parameters.
The four coefficients in \eqref{eq_potential}, $c_1$, $c_2$, $c_3$ and $c_4$,
are determined as follows:
$c_1 = 3(V_{+}-V_{-})/ (4 V_\mathrm{pos})$,
$c_2 = (V_{+}+V_{-})/V_\mathrm{pos}^2$,
$c_3 = -  ( V_{+} - V_{-})/(4 V_\mathrm{pos}^3)$,
$c_4 = -  ( V_{+} + V_{-})/(2 V_\mathrm{pos}^4)$.
Note that the another special point, $\widetilde{V}_\mathrm{pos}$,
is calculated as
$\widetilde{V}_\mathrm{pos} = - 3 V_\mathrm{pos} (V_{+} - V_{-}) / (8(V_{+} + V_{-}))$.
In addition, for later use,
the following abbreviation is introduced:
\begin{align}
a(x) = -c_1 - 2 c_2 x - 3 c_3 x^2 - 4 c_4 x^3,
\end{align}
and hence we can rewrite \eqref{eq_original_sde} as
\begin{align}
dx = a(x) \, dt + D \, dW.
\label{eq_original_sde_simple}
\end{align}

\subsection{Evaluated quantity}

In the present paper, an expectation value for a sigmoid-type function is evaluated.
For example, in escape problems, one may want to evaluate the probability with which the particle escapes from the local minimum in the double-well potential. 
That is, we have interests in the probability with which the particle is located at position $x > \alpha$, where $\alpha$ is a certain threshold.
To evaluate the probability, a step function should be used.
However, as we will see later, 
the proposed framework is not suitable for non-differentiable functions.
Hence, we here consider a kind of relaxation of the problem,
and a sigmoid function (logistic function) is used as the evaluated quantity:
\begin{align}
\sigma_h(x) = \frac{1}{1 + e^{-hx}},
\label{eq_sigmoid}
\end{align}
where $h$ is a parameter for determining the steepness of the function.
The aim of the present paper
is to evaluate the following expectation values for the sigmoid function;
\begin{align}
\mathbb{E}\left[ \sigma_h(X-\alpha) \right]_{T}
= \int_{-\infty}^{+\infty}
\sigma_h(x-\alpha) p(x,T) \, dx,
\label{eq_target}
\end{align}
where $p(x,T)$ is the probability density function of the system at time $T$.

When employing the conventional Monte Carlo simulations and obtaining $N$ samples, $x_1, \dots, x_N$, at time $T$, the expectation is approximately calculated as
\begin{align}
\mathbb{E}\left[ \sigma_h(X-\alpha) \right]_{T}
\simeq \frac{1}{N} \sum_{n=1}^{N} \sigma_h(x_n - \alpha).
\label{eq_target_MC}
\end{align}
Note that if we change the initial position of the particle at time $t=0$, the Monte Carlo simulations should be performed again using the different initial position. 
If one wants to know expectation values for the sigmoid function for various initial positions, it is necessary to perform the Monte Carlo simulations repeatedly; the repeated simulations need high computational costs.

\section{Derivation of the Corresponding Birth-Death Process}
\label{sec_derivation}

This section gives a main contribution of the present paper; the derivation of the birth-death process is demonstrated.
Since tedious calculations are needed to derive the final results, here only the outline of the derivation is shown; see Appendix A for the additional details of the derivation.

\subsection{Backward Fokker-Planck equation}

Firstly, using a traditional and conventional discussion of the backward Fokker-Planck equation, the original system in \eqref{eq_original_sde_simple} is connected to a partial differential equation.
Here, the following idea is employed in the present paper: \textit{let us consider the integrand as a new stochastic variable.}
This introduction of the dummy variable plays an important role from the viewpoint of computations, as we will see later.

In the present example, the following variable transformation is employed:
\begin{align}
x_1 = x, \qquad x_2 = \sigma_h(x-\alpha).
\label{eq_variable_transformation}
\end{align}
Note that the dummy variable $x_2$ is differentiable.
Hence, the following coupled stochastic differential equations are obtained by using the It{\^o} formula (as for the It{\^o} formula, see Ref.~\citen{Gardiner_book} for example):
\begin{align}
dx_1 =& \,a(x_1) dt + D \, dW, \\
dx_2 =& \, \left[ h x_2(1-x_2) a(x_1) 
+ \frac{h^2 D^2}{2} \left( x_2(1-x_2) - 2 x_2^2 (1-x_2) \right) \right] dt \nonumber \\
&+ h x_2(1-x_2) D \, dW.
\label{eq_coupled_sde}
\end{align}
From the definition in \eqref{eq_variable_transformation}, clearly the domain of $x_2$ is $(0,1)$.
For later use, the further variable transformation is employed here:
\begin{align}
z_1 = x_1, \qquad z_2 = 2x_2-1,
\end{align}
and hence $z_2 \in (-1,+1)$.
This variable transformation is used to connect the variable $z_2$ to the Legendre polynomials whose domain is $[-1,+1]$.
Then, the backward Fokker-Planck equation\cite{Gardiner_book,Risken_book} is obtained as 
\begin{align}
\partial_t g(\bm{z},t) = \mathcal{L}^\dagger g(\bm{z},t),
\end{align}
where
\begin{align}
\mathcal{L}^\dagger
=& a(z_1) \partial_1 
+ \Big\{ 
- \frac{h}{2} a(z_1) (z_2^2-1) \nonumber \\
&\qquad \qquad \,\, - \frac{h^2 D^2}{4} \left[ (z_2^2-1) - (z_2+1)(z_2^2-1) \right] 
\Big\} \partial_2 \nonumber \\
&+ \frac{D^2}{2} \partial_1^2 - h D^2 (z_2^2-1) \partial_1 \partial_2 
+ \frac{h^2 D^2}{8} (z_2^2-1)^2 \partial_2^2.
\label{eq_backward_Fokker_Planck}
\end{align}

Here, assume that $G(\bm{z})$ is a certain function.
Then, the backward Fokker-Planck equation or Feynman-Kac formula gives us the following fact\cite{Gardiner_book}; let $T$ be the time at which we want to calculate an expectation, and then the solution $g(x,t)$ of \eqref{eq_backward_Fokker_Planck} under the condition
\begin{align}
g(\bm{z},T) = G(\bm{z}),
\end{align}
is connected to the expectation for the original system as
\begin{align}
\left\langle G[\bm{z}(T)] | \bm{z}(0) = \bm{z}_0 \right\rangle
= g(\bm{z}_0,0),
\end{align}
where the left hand side means the conditional expectation of $G[\bm{z}(T)]$ under the initial condition $\bm{z}(0) = \bm{z}_0$.
That is, once we solve the backward Fokker-Planck equation and obtain $g(\bm{z},t=0)$, expectation values for \textit{arbitrary initial conditions} $\bm{z}_0$ (i.e., $\bm{x}_0$) are immediately obtained.
Here, our aim is to calculate the expectation value for $z_2$ (i.e, $x_2$), and hence we should set $G(\bm{z}) = z_2$.

The $g(\bm{z},t)$ should be solved backward in time.
To avoid confusion, we use the notation $\widetilde{g}(\bm{z},t) \equiv g(\bm{z}, T-t)$ below.

\subsection{Connection to birth-death process}

Secondly, the derived backward Fokker-Planck equation is converted to a birth-death process.
The conversion is achieved using a function expansion.
Although the similar technique has already been applied in the context of the dual stochastic process\cite{Ohkubo2019},
a further technique is employed here; because the domain of $x_2$ (i.e., $z_2$) is restricted, the Legendre polynomials are used in the function expansion.
The Legendre polynomials satisfy the following relations \cite{Koekoek_book}:
\begin{align}
\begin{cases}
\,\, (n+1) P_{n+1}(z_2) = (2n+1) z_2 P_n(z_2) - n P_{n-1}(z_2), \\
\,\, (z_2^2-1) \partial_2 P_n(z_2) = n z_2 P_n(z_2) - n P_{n-1}(z_2).
\end{cases}
\end{align}
See the factor $z_2^2-1$ in the left hand side in the second line; the same factor emerges in \eqref{eq_backward_Fokker_Planck}.
Hence, the Legendre polynomials would be adequate for this case, and actually, this gives numerical stability.

There are two variables $z_1$ and $z_2$ in $g(\bm{z},t)$.
As explained above, the range of $z_2$ is restricted in $(-1,+1)$; in contrast, $z_1$ takes arbitrary values in $\mathbb{R}$.
Hence, we employ the power-law-type basis functions for $z_1$, and the Legendre polynomials for $z_2$.
That is, the following basis expansion is used:
\begin{align}
\widetilde{g}(\bm{z},t) = \sum_{n_1=0}^\infty \sum_{n_2=0}^\infty P(n_1,n_2,t) z_1^{n_1} P_{n_2}(z_2),
\label{eq_basis_expansion}
\end{align}
where $\{P(n_1,n_2,t)\}$ are the expansion coefficients, and $P_{n}(z)$ is the Legendre polynomial with degree $n$.
Note that $P(n_1,n_2,t)$ is not a probability distribution because there is no probability conservation law in general.

Using the basis expansion in \eqref{eq_basis_expansion}, it is possible to derive the time-evolution equation for $\{P(n_1,n_2,t)\}$.
The derivation needs very complicated and tedious calculations; please see Appendix A for the steps of the derivation.
Appendix B gives the full form of the derived equations.

Although it is enough to use the derived equations to evaluate the expectations of the sigmoid function, it is possible to interpret the derived system as a birth-death process by considering adequate Feynman-Kac terms. 
In order to interpret the derived equations as a stochastic process, it would be suitable to use a slightly different kind of basis expansions as follows:
\begin{align}
\widetilde{g}(\bm{z},t) = \sum_{n_0=0}^\infty\sum_{n_1=0}^\infty \sum_{n_2=0}^\infty 
\widetilde{P}(n_0,n_1,n_2,t) (-1)^{n_0} z_1^{n_1} P_{n_2}(z_2),
\label{eq_basis_expansion_2}
\end{align}
where $n_0$ is newly introduced, which enables us to avoid the negative transition rates\cite{Ohkubo2013}.
Finally, we have the following $16$ reactions (for the interpretation, see Appendix B):
\begin{align}
&[n_1 \to n_1-2], \quad [n_1 \to n_1-1], \quad [n_1 \to n_1 + 1], \nonumber \\
&[n_1 \to n_1+2], \quad [n_1 \to n_1-1 \, \&\,  n_2 \to n_2-1], \nonumber \\
&[n_1 \to n_1-1 \, \&\,  n_2 \to n_2+1], \quad [n_1 \to n_1+1 \, \& \, n_2 \to n_2-1], \nonumber \\
&[n_1 \to n_1+1 \, \& \, n_2 \to n_2+1], \quad [n_1 \to n_1+2 \, \&\,  n_2 \to n_2-1], \nonumber \\
&[n_1 \to n_1+2 \, \&\,  n_2 \to n_2+1], \quad [n_1 \to n_1+3 \, \& \, n_2 \to n_2-1], \nonumber \\
&[n_1 \to n_1+3 \, \&\,  n_2 \to n_2+1], \quad [n_2 \to n_2-2], \nonumber \\
&[n_2 \to n_2-1], \quad [n_2 \to n_2+1], \quad [n_2 \to n_2+2].
\end{align}
Note that if the coefficient (transition rate) is negative, the reaction of $[n_0 \to n_0+1]$ should be added for the corresponding reaction.

\subsection{Some comments for the proposed framework}

As demonstrated in the next section, the proposed method gives us some merits from the computational point of view.

The first merit is related to numerical solvers of the system.
If one wants to solve the backward Fokker-Planck equation directly, the discretization for time and space is needed.
The basis expansion enables us to solve the problem as the coupled ordinary differential equations;
although a finite cut-off is needed, numerical time-evolution methods, such as the Runge-Kutta method, are available without using space discretization.
Furthermore, when we have a larger number of variables, we will suffer from the curse of dimensionality.
However, since we interpret the system as a birth-death process, it is possible to use the Monte Carlo methods for birth-death processes (for example, the famous Gillespie algorithm \cite{Gillespie1977}) instead of the deterministic numerical methods.

The second merit is related to the parameters in the expectation.
As explained above, it is possible to calculate expectations for various initial conditions using the proposed method.
Additionally, as demonstrated in the next section, the introduction of the dummy variable enables us
to evaluate expectations for various parameter $\alpha$ in the sigmoid function 
\textit{all at once}.

\section{Numerical Checks}
\label{sec_numerical_results}

In this section, we check numerically whether the proposed framework works well or not.
For the parameters of the double-well potential, $V_\mathrm{pos} = 1$, $V_{-} = -0.1$, $V_{+} = -2.0$ are used.
Note that these parameters give $\widetilde{V}_\mathrm{pos} \simeq -0.339$.
Additionally, the parameter for the sigmoid function is $h = 5$, and $D = 0.5$ is used as the diffusion constant.

Here, the direct numerical time-evolution of the obtained process, i.e., the coupled ordinary differential equations in Appendix B, is employed.
In principle, there are infinite numbers of equations, and then we should use a cutoff; here the cutoff with $0 \le n_1 \le 50$ and $0 \le n_2 \le 50$ is used.
The 4th order Runge-Kutta method with $\Delta t = 10^{-5}$ is employed, and the expectation values of the sigmoid function at $T = 0.1$ are considered here.

\begin{figure}
\begin{center}
\includegraphics[width=70mm]{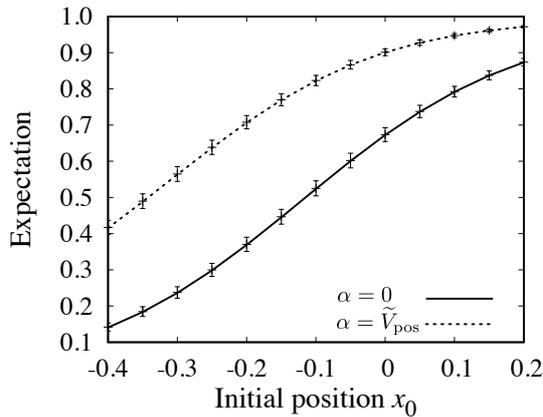}
\end{center}
\caption{\label{fig_results} 
Numerical evaluation results.
The solid and dotted curves correspond to the evaluation results via the proposed method, and the points with error bars are obtained from the direct Monte Carlo samplings.
Note that both curves are obtained only from a numerical solution for the derived process; the values for different initial positions $x_0$ and different thresholds $\alpha$ are obtained from the same $\{P(n_1,n_2,T)\}$.
}
\end{figure}

Using the Mac Book (1.4GHz, Intel Core i7), the time-evolution of the derived process in Appendix B was finished less than $2$ minutes.
Using the proposed framework in the present paper, only one solution $\{P(n_1,n_2,T)\}$ is enough to evaluate
the expectation in \eqref{eq_target} for \textit{various} initial conditions.
Actually, we can draw a curve in Fig.~\ref{fig_results} using only one solution $\{P(n_1,n_2,T)\}$.
The calculation for each initial position is very simple and the computational costs are low; we have $13$ points in $[-0.4, 0.2]$ to depict a curve in Fig.~\ref{fig_results}, and a naive code written in Python needs less than $0.5$ seconds for the calculation of these $13$ points.

For comparison, direct Monte Carlo samplings for the original stochastic differential equation were also performed.
Using the Euler-Maruyama method with $\Delta t = 10^{-4}$ and setting the sampling number $N$ as $100$, the expectation in \eqref{eq_target_MC} was evaluated.
Additionally, $100$ trials were repeated to obtain each error bar in Fig.~\ref{fig_results}.
That is, for one point in Fig.~\ref{fig_results}, $10000$ samples were used.
To obtain $13$ points (with error bars) in a curve, a naive code written in C needed about $15$ seconds.
We need more computational costs in the direct Monte Carlo samplings when we want to have less error bars and to have results with other initial positions.

Furthermore, although there are two curves, i.e., $\alpha = 0$ and $\alpha = \widetilde{V}_\mathrm{pos}$ cases in Fig.~\ref{fig_results}, only one solution $\{P(n_1,n_2,T)\}$ is enough to depict the both curves.
The only difference is the change of the value $z_2$ in \eqref{eq_basis_expansion}.
This is also one of the computational merits.

\section{Concluding Remarks}
\label{sec_concluding_remarks}

In the present paper, the introduction of dummy variables, the usage of the It\^o formula, and the basis expansions are employed for the backward Fokker-Planck equation.
These techniques enable us to derive the corresponding birth-death process.
Additionally, it is also clarified that the proposed method has some merits from the computational point of view.

Here, some issues of the proposed framework should be mentioned.
One of the issues stems from the basis function.
The proposed method uses the basis expansions, and then the domains of the variables are restricted largely; wider domains need larger cutoffs for the coefficients in the basis function.
Those lead to higher computational costs for the time-evolution of the birth-death process and to larger memories to store the coefficients.
The second issue is the restriction for the integrands; it is easy to see from the derivation 
that the integrands should be differentiable functions.
No ideas have been proposed so far to deal with non-differentiable functions.
Additionally, it is not clear what kind of basis functions should be used.
In the present paper, the Legendre polynomials are used for the sigmoid function.
It is possible to use other orthogonal polynomials; the choice of suitable polynomials from the practical viewpoint is also a remaining future work.

As shown in the numerical experiments, the proposed method could open up an efficient numerical technique
for evaluating expectations; that is, the following numerical procedures were employed:
\begin{enumerate}
\item[(i)] A numerical evaluation for the derived birth-death process is performed \textit{only once} in advance.
\item[(ii)] Employing the pre-calculated results, it is possible to evaluate the expectation values for the sigmoid-type integrand for \textit{various} initial conditions.
\item[(iii)]
Also, the \textit{same} pre-calculated results are employed in order to evaluate the expectation values for \textit{various} center positions of the sigmoid-type integrand.
\end{enumerate}
Note that only \textit{one} numerical evaluation of the birth-death process gives the expectation values of a certain statistic (here, the sigmoid-type) for \textit{various} conditions, which is achieved by the separation of the pre-calculation and the run-time calculation.
One of the possible application fields are control theory; for example, optimal controls in stochastic systems were discussed in terms of path-integrals \cite{Kappen2005,Williams2017,Okada2018}, and the expectation for Gaussian functions were evaluated in the problem.
We need further numerical techniques to apply the proposed method in the present paper to the optimal controls, and the application is now under investigation; we will need not only the backward Fokker-Planck equation but also the Feynman-Kac formula to treat the control problems.
However, the separation of the calculations for the birth-death processes and the run-time calculations for the original system could be efficient and useful in practice.
I hope that the present paper stimulates various researchers not only in physics but also in other interdisciplinary research communities.

\vspace{1em}
\begin{acknowledgment}

\acknowledgment

This work was supported by JST, PRESTO Grant Number JPMJPR18M4, Japan.

%For enveironments for acknowledgment(s) are available: \verb|acknowledgment|, \verb|acknowledgments|, \verb|acknowledgment|, and \verb|acknowledgments|.

\end{acknowledgment}

%\appndix
%\section{}
%Use the \verb|\appendix| command if you need an appendix(es). The \verb|\section| command should follow even though there is no title for the appendix (see above in the source of this file).
%For authurs of Invited Review Papers, the \verb|profile| command si prepared for the author(s)' profile. A simple example is shown below.

\appendix

\section{Some demonstrations of the derivation}

Using the basis expansions, coupled ordinary differential equations for the coefficients $\{P(n_1,n_2,t)\}$
are derived instead of the partial differential equation for $g(\bm{z},t)$.
The derivation is simple and straightforward, but tedious.
The full form is shown in Appendix B, and here only a demonstration of the derivation is given for a term in \eqref{eq_backward_Fokker_Planck}.
There is a term 
\begin{align}
-\frac{h}{2} (-2 c_2 z_1) (z_2^2 -1) \partial_2
\end{align}
in the first curly bracket in \eqref{eq_backward_Fokker_Planck}.
Note that the abbreviation of $a(z_1) = -c_1 -2 c_2 z_1 - 3 c_3 z_1^2 - 4 c_4 z_1^3$; the second term in this abbreviation is considered here.
Then, employing the basis expansion in \eqref{eq_basis_expansion}, we have
\begin{align}
&-\frac{h}{2} (-2 c_2 z_1) (z_2^2 -1) \partial_2 
\sum_{n_1=0}^{\infty} \sum_{n_2=0}^{\infty} P(n_1,n_2,t) z_1^{n_1} P_{n_2}(z_2)
\nonumber \\
&=
h c_2 (z_2^2 -1) \partial_2 
\sum_{n_1=0}^{\infty} \sum_{n_2=0}^{\infty} P(n_1,n_2,t) z_1^{n_1+1} P_{n_2}(z_2) \nonumber \\
&=
h c_2
\sum_{n_1=1}^{\infty} \sum_{n_2=1}^{\infty} 
P(n_1-1,n_2,t) z_1^{n_1} n_2 z_2 P_{n_2}(z_2) \nonumber \\
&\quad - h c_2
\sum_{n_1=1}^{\infty} \sum_{n_2=1}^{\infty} 
P(n_1-1,n_2,t) z_1^{n_1} n_2 P_{n_2-1}(z_2)
\nonumber \\
&= 
h c_2
\sum_{n_1=1}^{\infty} \sum_{n_2=1}^{\infty} 
P(n_1-1,n_2,t) z_1^{n_1} n_2 \frac{n_2+1}{2n_2+1} P_{n_2+1}(z_2) \nonumber \\
&\quad + h c_2
\sum_{n_1=1}^{\infty} \sum_{n_2=1}^{\infty} 
P(n_1-1,n_2,t) z_1^{n_1} n_2 \frac{n_2}{2n_2+1} P_{n_2-1}(z_2) \nonumber \\
&\quad - h c_2
\sum_{n_1=1}^{\infty} \sum_{n_2=1}^{\infty} 
P(n_1-1,n_2,t) z_1^{n_1} n_2 P_{n_2-1}(z_2)
\nonumber \\
&= 
h c_2
\sum_{n_1=1}^{\infty} \sum_{n_2=1}^{\infty} 
P(n_1-1,n_2-1,t) z_1^{n_1} \frac{(n_2-1)n_2}{2n_2-1} P_{n_2}(z_2) \nonumber \\
&\quad + h c_2
\sum_{n_1=1}^{\infty} \sum_{n_2=0}^{\infty} 
P(n_1-1,n_2+1,t) z_1^{n_1} \frac{(n_2+1)^2}{2n_2+3} P_{n_2}(z_2) \nonumber \\
&\quad - h c_2
\sum_{n_1=1}^{\infty} \sum_{n_2=1}^{\infty} 
P(n_1-1,n_2+1,t) z_1^{n_1} (n_2+1) P_{n_2}(z_2),
\end{align}
and hence we obtain the following terms
in the coupled ordinary differential equations for $P(n_1,n_2,t)$:
\begin{align*}
&h c_2 \frac{(n_2-1)n_2}{2n_2-1} P(n_1-1,n_2-1,t) \nonumber \\
&+ h c_2 \frac{(n_2+1)^2}{2n_2+3} P(n_1-1,n_2+1,t) \nonumber \\
&- h c_2 (n_2+1) P(n_1-1,n_2+1,t).
\end{align*}
As for the full form, see Appendix B.

\section{Full form of the time-evolution equations for the dual process}

After similar calculations shown in Appendix A, we finally obtain the following coupled ordinary differential equation:
\begin{align}
&\frac{\partial}{\partial t} P(n_1, n_2, t) \nonumber \\
=& 
\frac{D^2}{2} (n_1+1)(n_1+2) P(n_1+2,n_2,t) \nonumber \\
&- c_1(n_1+1) P(n_1+1,n_2,t) \nonumber \\
&- 3c_3 (n_1-1) P(n_1-1,n_2,t) \nonumber \\
&- 4c_4(n_1-2)P(n_1-2,n_2,t) \nonumber \\
&
- \frac{h D^2}{2} \Bigg[ \frac{(n_1+1)(n_2+1)^2}{2n_2+3} \nonumber \\
&\qquad + (n_1+1)(n_2+1) \Bigg] P(n_1+1,n_2+1,t)  \nonumber \\
&- \frac{h D^2}{2} \frac{(n_1+1)(n_2-1)n_2}{2n_2-1} P(n_1+1,n_2-1,t) \nonumber \\
&+ h c_2 
\left[ \frac{(n_2+1)^2}{2n_2+3} - (n_2+1) \right] P(n_1-1,n_2+1,t) \nonumber \\
& + h c_2 \frac{(n_2-1)n_2}{2n_2-1} P(n_1-1,n_2-1,t) \nonumber \\
&+ \frac{3 h c_3}{2}
\left[ \frac{(n_2+1)^2}{2n_2+3} - (n_2+1) \right] P(n_1-2,n_2+1,t) \nonumber \\
&+ \frac{3 h c_3}{2} \frac{(n_2-1)n_2}{2n_2-1} P(n_1-2,n_2-1,t)\nonumber \\
&+ 2 h c_4 \left[ \frac{(n_2+1)^2}{2n_2+3} - (n_2+1) \right] P(n_1-3,n_2+1,t) \nonumber \\
&+ 2 h c_4 \frac{(n_2-1)n_2}{2n_2-1} P(n_1-3,n_2-1,t) \nonumber \\
&+ \frac{h^2 D^2}{8}
\Bigg[ 
\frac{(n_2+2)^2 (n_2+1)}{(2n_2+5)(2n_2+3)} 
+ \frac{(n_2+2)^3(n_2+1)}{(2n_2+5)(2n_2+3)} \nonumber \\
&\qquad - \frac{(n_2+2)^2(n_2+1)}{2n_2+3} 
- \frac{2(n_2+2)^2 (n_2+1)}{(2n_2+5)(2n_2+3)} \nonumber \\
&\qquad - \frac{(n_2+2)(n_2+1)^2}{2n_2+3} 
+ (n_2+2)(n_2+1) \nonumber \\
&\qquad + \frac{2(n_2+2)(n_2+1)}{2n_2+3}
+ \frac{2(n_2+2)^2(n_2+1)}{(2n_2+5)(2n_2+3)} \nonumber \\
&\qquad - \frac{2(n_2+2)(n_2+1)}{2n_2+3}
\Bigg] P(n_1,n_2+2,t) \nonumber \\
&+ \frac{h c_1}{2} 
\left[ \frac{(n_2+1)^2}{2n_2+3} -  (n_2+1) \right] P(n_1,n_2+1,t) \nonumber \\
&+ \Bigg[
\frac{h c_1}{2} \frac{(n_2-1)n_2}{2n_2-1}
- \frac{h^2 D^2}{4} \frac{n_2(n_2-1)}{2n_2-1} \nonumber \\
&\qquad + \frac{h^2 D^2}{4} \frac{(n_2-1)n_2}{2n_2-1}
\Bigg] P(n_1,n_2-1,t) \nonumber \\
&+ \frac{h^2 D^2}{8} 
\Bigg[ 
\frac{(n_2-2)(n_2-1)n_2}{(2n_1-3)(2n_2-1)} 
+ \frac{(n_2-2)^2 (n_2-1) n_2}{(2n_2-3)(2n_2-1)} \nonumber \\
&\qquad - \frac{2(n_2-2)(n_2-1)n_2}{(2n_2-3)(2n_2-1)} \nonumber \\
&\qquad + \frac{2(n_2-2)(n_2-1)n_2}{(2n_2-3)(2n_2-1)}
\Bigg] P(n_1,n_2-2,t) \nonumber \\
&+ \Bigg\{
- 2c_2 n_1 
+ \frac{h^2 D^2}{8} \Bigg[  
\frac{n_2 (n_2+1)^2}{(2n_1+1)(2n_2+3)}  \nonumber \\
& \qquad +  \frac{n_2^3}{(2n_1+1)(2n_2-1)} 
- n_2 
+ \frac{n_2^2 (n_2+1)^2}{(2n_2+1)(2n_2+3)} \nonumber \\
&\qquad + \frac{n_2^4}{(2n_2+1)(2n_2-1)}
- \frac{n_2^3}{2n_2-1}
- \frac{n_2^2 (n_2-1)}{2n_2-1}   \nonumber \\
&\qquad - \frac{2n_2(n_2+1)^2}{(2n_2+1)(2n_2+3)}
- \frac{2n_2^3}{(2n_2+1)(2n_2-1)} \nonumber \\
&\qquad + \frac{2n_2^2}{2n_2-1} 
+ \frac{2n_2(n_2+1)^2}{(2n_2+1)(2n_2+3)} \nonumber \\
&\qquad + \frac{2n_2^3}{(2n_2+1)(2n_2-1)} 
- \frac{2n_2^2}{2n_2-1}
\Bigg]
\Bigg\} P(n_1,n_2,t).
\label{eq_appendix_full_equations}
\end{align}

Note that, as shown in Ref.~\citen{Ohkubo2013}, it is possible to convert the above system of the coupled ordinary differential equations to the corresponding \textit{stochastic process}.
Here, only some demonstrations are shown.
For example, the first term in the right hand side corresponds to the death process
\begin{align}
n_1 \to n_1-2 \quad \textrm{at rate $D^2 n_1(n_1-1)/2$},
\end{align}
because $P(n_1+2,n_2,t)$ contributes to the time-derivative of $P(n_1,n_2,t)$; two particles are decreased by this contribution.
We need an additional term to guarantee the probability conservation law.
As for the second term, we have
\begin{align}
\begin{cases}
n_1 + 1 \to n_1\\
n_0 \to n_0 +1
\end{cases}
\quad \textrm{at rate $c_1 n_1$},
\end{align}
where $n_0$ is introduced to avoid the negative transition problem \cite{Ohkubo2013}.
In the present paper, the number of variables is small enough, and we can perform the direct time-integration of \eqref{eq_appendix_full_equations}.
Hence, the conversion was not necessary.

%\begin{verbatim}
%\profile{Taro Butsuri}{was born in Tokyo, Japan in 1965. ...}
%\end{verbatim}

\end{document}